\documentclass[numbers,webpdf,imaiai]{ima-authoring-template}
\graphicspath{{Fig/}}
\usepackage{epstopdf}
\usepackage{bm}
\theoremstyle{thmstyleone}%
\newtheorem{theorem}{Theorem}
\theoremstyle{thmstyletwo}%
\newtheorem{remark}{Remark}%
\theoremstyle{thmstylethree}%

\numberwithin{equation}{section}
\newcommand{\FDK}{\mathcal C^\dagger_{\mbox{\tiny FDK}}}

                    {$\blacksquare$\vspace*{7pt}} 

\def\R{{\mathbb R}}

\def\ssu{\mathrm{sing\text{-}supp}}

\def\dim{\mathrm{dim}}

\def\Om{\Omega}
\def\om{\omega}

\def\f{\frac}
\def\p{\partial}
\def\q{\quad}

\def\bi{{\mathbf i}}

\def\x{\boldsymbol{x}}
\def\y{{\boldsymbol y}}
\def\z{{\boldsymbol z}}
\def\A{{\mathbf A}}

\def\I{{\mathbf I}}

\def\W{{\mathbf W}}

\def\mH{{\mathcal H}}
\def\mI{{\mathcal I}}

\def\mL{{\mathcal L}}

\def\mR{{\mathcal R}}
\def\mS{{\mathcal S}}

\def\msC{{\mathscr C}}

\def\msF{{\mathscr F}}

\def\mmu{\boldsymbol{\mu}}

\def\bi{\begin{itemize}} \def\ei{\end{itemize}}
\def\be{\begin{eqnarray*}}
\def\ee{\end{eqnarray*}}

\def\etal{{\it et al }}

\def\0{{\mathbf 0}}

\newcommand{\beq}{\begin{equation}}
\newcommand{\eeq}{\end{equation}}

\def\bth{\boldsymbol{\theta}}

\def\eref#1{(\ref{#1})}

\newcommand{\eps}{\varepsilon}
\def\x{\mathbf{x}}
\def\y{\mathbf{y}}
\def\z{\mathbf{z}}

\def\XXint#1#2#3{{\setbox0=\hbox{$#1{#2#3}{\int}$ }
\vcenter{\hbox{$#2#3$ }}\kern-.55\wd0}}

\begin{document}

\copyrightyear{2023}
\firstpage{1}

\title[Nonlinear ill-posed problem in low-dose dental CBCT]{Nonlinear ill-posed problem in low-dose dental cone-beam computed tomography\footnote{Dedicated to the memory of Prof. Escauriaza Luis.}}

\authormark{H. S. Park et al.}
\author{\textsc{Hyoung Suk Park}
\address{\orgname{National Institute for Mathematical Sciences}, \orgaddress{\street{Daejeon}, \country{Republic of Korea}}}}
\author{\textsc{Chang Min Hyun$^\star$, Jin Keun Seo} \address{\orgdiv{School of Mathematics and Computing (Computational Science and Engineering)}, \\ \orgname{Yonsei University}, \orgaddress{\street{Seoul}, \country{Republic of Korea}}}}
\corresp[$\star$]{Corresponding author : \href{email:chammyhyun@yonsei.ac.kr}{chammyhyun@yonsei.ac.kr}}
\received{Date}{0}{Year}
\revised{Date}{0}{Year}
\accepted{Date}{0}{Year}


\abstract{This paper describes the mathematical structure of the ill-posed nonlinear inverse problem of low-dose dental cone-beam computed tomography (CBCT) and explains the advantages of a deep learning-based approach to the reconstruction of computed tomography images over conventional regularization methods. This paper explains the underlying reasons why dental CBCT is more ill-posed than standard computed tomography. Despite this severe ill-posedness, the demand for dental CBCT systems is rapidly growing because of their cost competitiveness and low radiation dose. We then describe the limitations of existing methods in the accurate restoration of the morphological structures of teeth using dental CBCT data severely damaged by metal implants. We further discuss the usefulness of panoramic images generated from CBCT data for accurate tooth segmentation. We also discuss the possibility of utilizing radiation-free intra-oral scan data as prior information in CBCT image reconstruction to compensate for the damage to data caused by metal implants.}

\keywords{cone-beam computed tomography; ill-posed inverse problem; deep learning; metal artifact reduction.}

\def\Et{{\mbox{{\tiny $E$}}}}
\maketitle

\section{Introduction}
Computed tomography (CT) is an established diagnostic imaging tool that  produces cross-sectional images (i.e., slices of anatomy) using projection data (denoted by $\text{P}$), which consist of a series of X-ray images taken from various angles in the human body. It uses the X-ray ionizing radiation (electromagnetic waves with wavelengths ranging from $10^{-8}$ to $10^{-12}$ m) discovered by Wilhelm R\"ontgen, who was awarded the first Nobel Prize in Physics in 1901 \cite{Haase1997}. In medical CT, polychromatic  X-ray beams pass through the body and to a two-dimensional array of detectors that acquires the projection data (i.e., X-ray images).

The relationship between projection data $\text{P}$ and CT image $\mu$ (assigning an X-ray attenuation coefficient)  is given by Lambert-Beer's law \cite{Beer1852,Lambert1892}, which, in a two-dimensional (2D) parallel beam CT model, is
\begin{align}\label{P_theta}
\text{P}^{\sharp}(\theta,s)=-\ln\left(\int\eta(E)
\exp\big\{-[\mR(\mu_\Et)](\theta,s)\big\}dE\right),
\end{align}
where $\text{P}^{\sharp}(\theta,s)$ denotes the projection data of the 2D parallel beam CT at projection angle $\theta\in[0,2\pi)$ and detector position $s$,  $\mu_\Et$ denotes the attenuation coefficient distribution at photon energy level $E$,  $\eta(E)$ represents the fractional energy at $E$ \cite{Herman1983,Poludniowski2009}, and $\mR(\mu_\Et)$ is the Radon transform of $\mu_\Et$. Although CT theory began based on this parallel beam model, parallel CT is not used in clinical practice. However, parallel CT is very similar to fan-beam multi-detector CT (MDCT) from the mathematical point of view in CT image reconstruction, and MDCT is widely used in clinical practice.

Conventional CT image reconstruction is based on the ideal monochromatic assumption; $\eta(E)=\delta(E-E_0)$ in (\ref{P_theta}), where $\delta$ is the Dirac delta function and $E_0$ is a fixed energy. Under this monochromatic assumption, the inverse problem  (\ref{P_theta}) becomes the well-posed linear inverse problem
$\text{P}^{\sharp}=\mR \mu$, where $\mu=\mu_{\Et_0}$. In 1917, Johann Radon \cite{Radon1917} found that a CT image $\mu$ could be reconstructed from $\text{P}^{\sharp}$ (X-ray images at all directions). In the late 1960s, the first clinical CT scanner was developed by Hounsfield \cite{Hounsfield1973}, and  Hounsfield and Cormack \cite{Cormack1963} shared the Nobel Prize for Medicine in 1979. The CT image reconstruction algorithm currently used for clinical CT (or MDCT) is based on the filtered back-projection (FBP) algorithm \cite{Bracewell1967}, built under the ideal assumption that $\text{P}^{\sharp}$ is in the range of the Radon transform \cite{Radon2005}.

However, the actual model of CT is nonlinear due to the polychromatic nature of the incident X-ray beam. Hence, an inconsistency exists in the mathematical model $\text{P}^{\sharp}=\mR \mu$ because of the interaction between the applied X-ray beam and tissues. FBP ignores the nonlinear structure of $P$ produced in terms of the change in tissue property $\mu$ with respect to $E$ along the fractional energy distribution $\eta(E)$ in \eqref{P_theta}. The $\mu_\Et$ of high attenuation materials such as metals varies significantly with $E$, and hence the presence of metals in the field-of-view (FOV) of clinical CT produces a serious discrepancy in the model $\text{P}^{\sharp}=\mR \mu$, resulting in serious streaking artifacts associated with the metal geometry when the FBP algorithm is applied. Metal artifact reduction (MAR) in CT  is becoming increasingly important as the number of older adults with artificial prostheses and metallic implants increases rapidly. Unfortunately, although numerous MAR methods have been suggested  over the past 40 years \cite{Alvarez1976,Lehmann1981,Yu2012}, MAR remains a difficult problem because of the nonlinear structure of the inverse problem (\ref{P_theta}) associated with $\eta(E)$ \cite{Park2017}.

Although MDCT is known as the most accurate and reliable imaging technique, it is rarely used in small dental clinics because of its high equipment cost and the large space required for its use. By contrast, dental cone-beam CT (CBCT), an alternative to MDCT, is increasingly being used in dental clinics. Dental CBCT is an important element of digital dentistry \cite{elnagar2020digital} and is used in various dental fields such as implant/prosthetics, oral and maxillofacial surgery, and orthodontic treatment.  Most dental CBCT devices are designed to allow the patient to be scanned while sitting or standing, requiring less space in the dental office. Moreover, most dental CBCT devices are designed to reduce the radiation dose by limiting the scan's FOV \cite{Zhen2013}.

\begin{figure}[ht]
	\centering
	\includegraphics[width=1\textwidth]{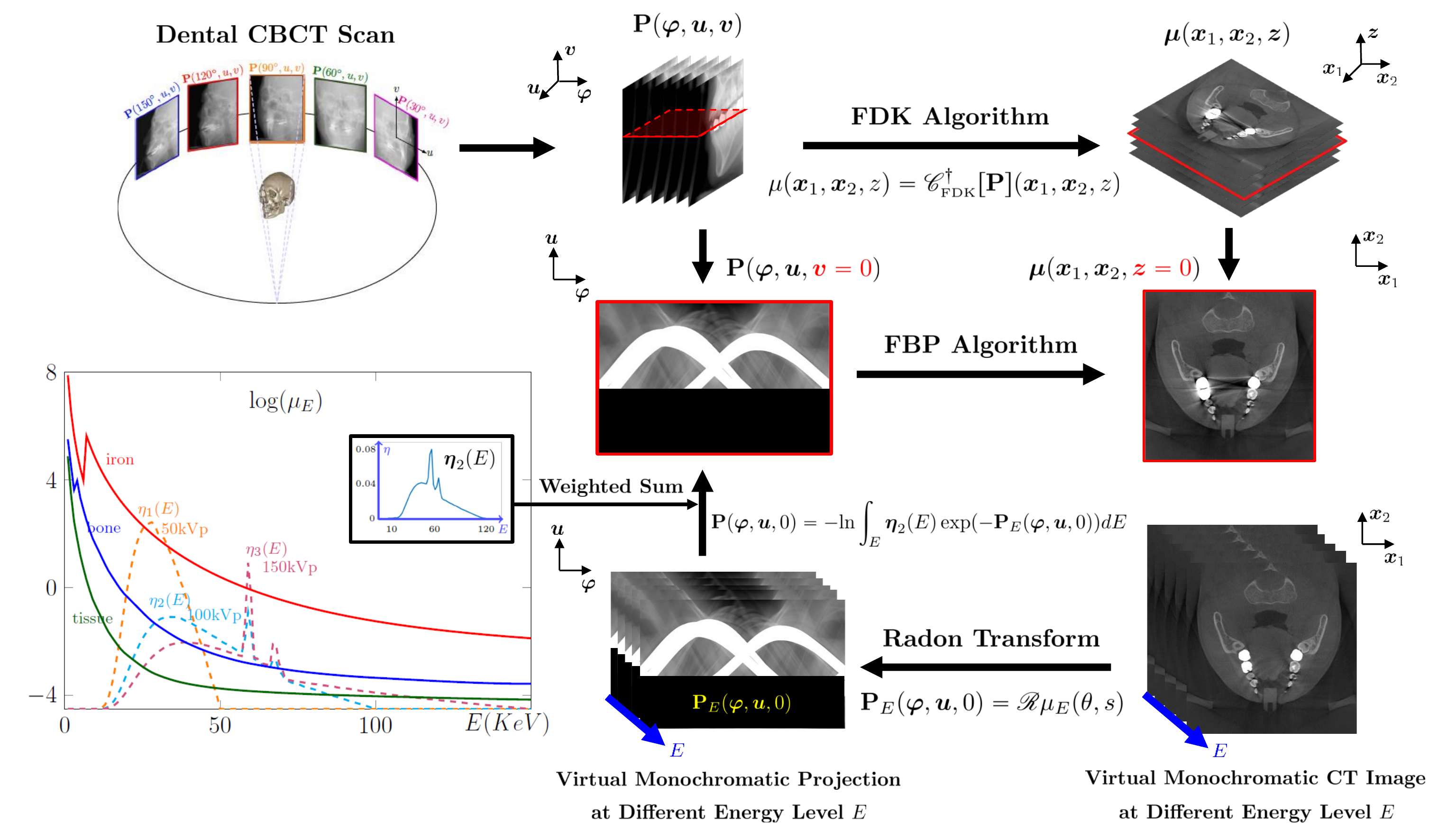}
	\caption{Nonlinear inverse problem in polychromatic X-ray CT image reconstruction. The polychromatic data $\text{P}$ can be viewed as the spectrum-weighted sum of monochromatic data $\text{P}_\Et$, which leads to serious mismatch between the real projection data and mathematical model in the presence of metallic implants. Due to the inherent nature of FBP (i.e., the pseudo inverse of the Radon transform), the local inconsistency of $\text{P}(\varphi, u,0)$ generates severe global artifacts in the reconstructed image $\mu(\x,0)$ (the right middle image), which appear as streaking and shading artifacts. }
	\label{CBCTForwardRecon}
\end{figure}

\begin{figure}[ht]
	\centering
	\includegraphics[width=1\textwidth]{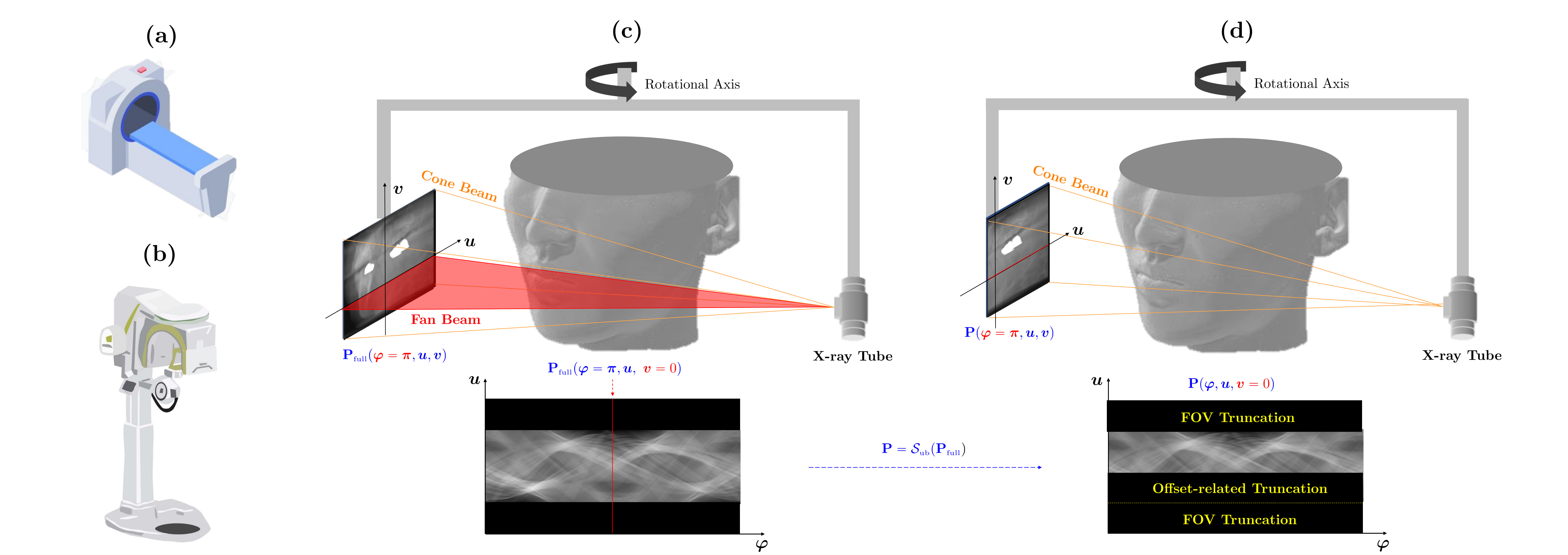}
	\caption{MDCT and CBCT geometry. (a) MDCT: Conventional CT acquires projection data using a fan-shape X-ray beam that moves in a spiral. (b) Dental CBCT uses a rectangular cone-shape beam that rotates once. (c) CBCT scan with a full detector. (d) CBCT-scan with an offset detector. Most dental CBCT machines use an offset detector to acquire only half of the extended FOV with a single projection.}
	\label{cone-beam}
\end{figure}

\begin{figure}
	\centering
	\includegraphics[width=1\textwidth]{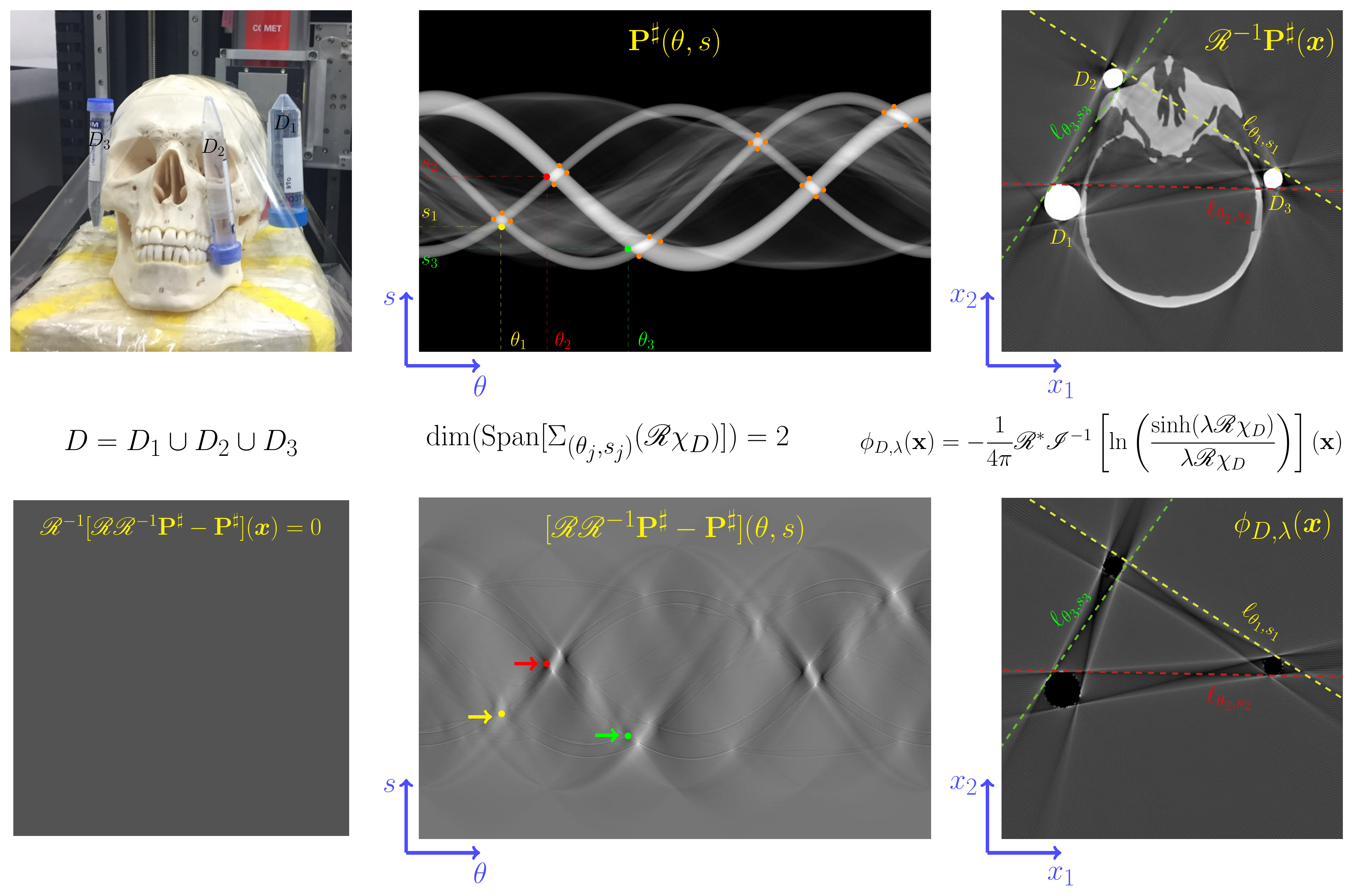}
	\caption{Metal-induced streaking artifacts. The beam hardening corrector $\phi_{D,\lambda}$ is a function of $D$ (metal geometry) and the control parameter $\lambda$ associated with all energy dependent factors including attenuation coefficients and spectrum of X-ray source. Here, $\mathcal I^{-1}$ is the Riesz potential of degree -1.}
	\label{metalartifact}
\end{figure}

Currently, dental CBCT is being developed in the direction of providing high-resolution images (like those of MDCT), while optimizing data collection in terms of low invasiveness and cost-effectiveness. This willingness leads to a highly ill-posed inverse problem in the sense of Hadamard \cite{Hadamard1902} as follows:
\begin{itemize}
  \item{Non-existence:}  The existence of a solution is not guaranteed because most of the measured sinogram data do not match a linear model. Given the polychromatic nature of the incident X-ray beam, discrepancies may exist between the actual projection data and the mathematical model, as illustrated in Fig. \ref{CBCTForwardRecon}.
  \item{Non-uniqueness:} Low-dose dental CBCT can be viewed as an underdetermined problem (fewer equations than unknowns) because of its intention to provide high-resolution images with as few data as possible, as illustrated in Fig. \ref{cone-beam} (b).
  \item{Instability:} A local inconsistency in the projection data can generate severe global streaking artifacts, as shown in Fig. \ref{metalartifact}.
\end{itemize}
As the demand to reduce the X-ray dose and maintenance costs become increasingly severe, the inverse problem of CBCT becomes increasingly ill-posed.

This paper describes the mathematical structure of the ill-posed inverse problem of low-dose dental CBCT and explains the advantages of a deep learning-based approach for CT image reconstruction over conventional regularization methods.

Solving the ill-posed problem of low-dose dental CBCT requires the use of an image prior for the expected images. Various regularization techniques for constraining the solution have been proposed, such as total variation (TV) \cite{Osher2005,Sidky2011}, sparse representations \cite{Donoho2006,Candes2005,Kudo2013} and deep learning-based models \cite{Jin2017,Bora2017,Ulyanov2020}. Although these methods achieved remarkable performance in reducing noise, they still have a limited ability to handle metal-induced artifacts in the dental CBCT environment, where metal artifacts are common. Handling metal-induced artifacts in dental CBCT is difficult because its global structure is not only non-linearly influenced by local metal geometry, but is also entangled with complex factors associated with metal-bone and metal-teeth interactions, FOV truncation, offset detector acquisition, scattering, nonlinear partial volume effects, and other factors \cite{Barrett2004,Bayaraa2020,Hsieh2002}. This paper describes recent complementary approaches to acquiring concrete image priors using generated panoramic images and a radiation-free oral scanner.

\section{Mathematical framework of low-dose dental CBCT}
\begin{figure}[t!]
	\centering
	\includegraphics[width=1\textwidth]{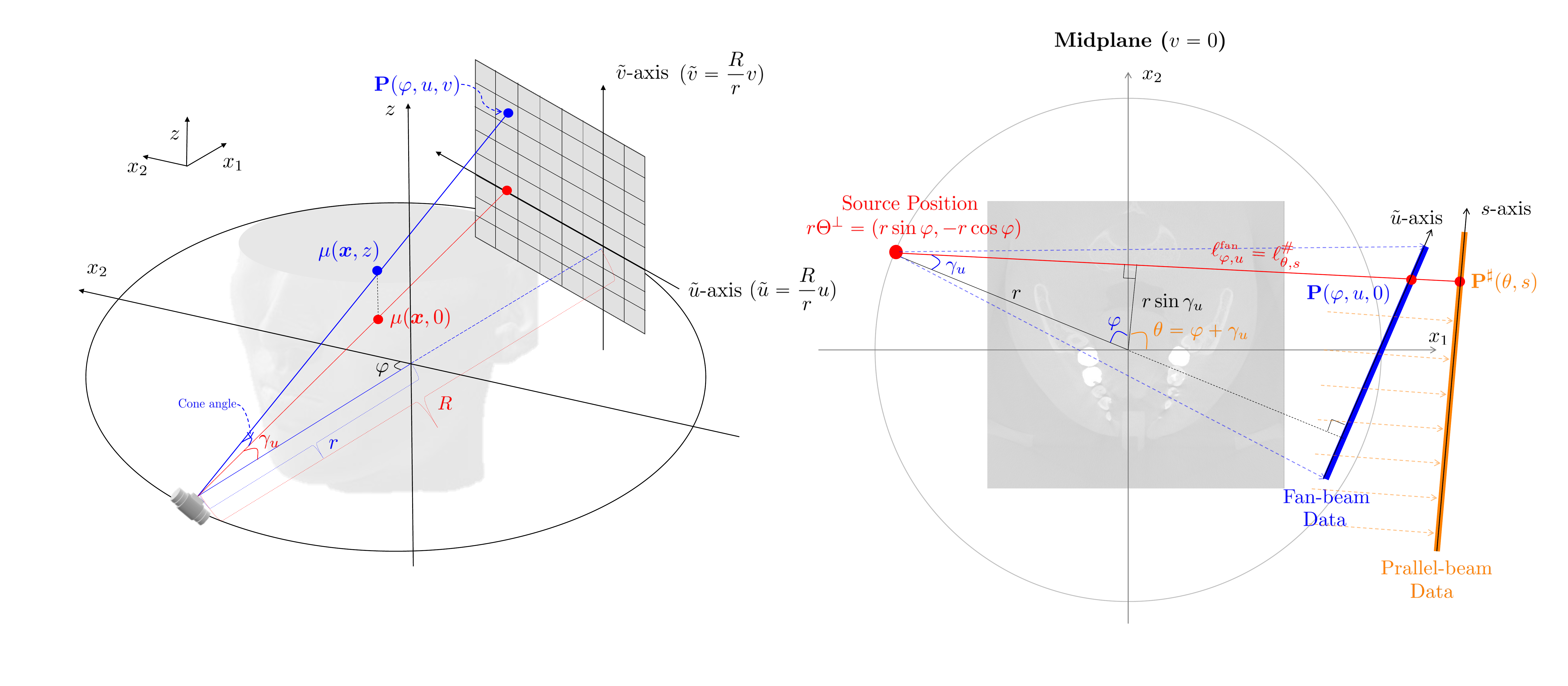}
	\caption{(Left) Cone-beam projection geometry and (Right) the relation between parallel-beam data $\mbox{\textbf{P}}^\sharp(\theta,s)$ and cone-beam data at the midplane (fan-beam data) $\mbox{\textbf{P}}(\varphi,u,0)$ . When representing the projection data $\mbox{\textbf{P}}(\varphi,u,v)$, for sake of the simplicity, the detector plane is pretended to lie on axis of rotation. The actual physical location corresponding to the position $(u,v)$ is $\frac{R}{r}(u,v)$.}
	\label{CBCTGeometry}
\end{figure}

\subsection{Inverse problem}
The mathematical model of low-dose dental CBCT (planar detector) can be described using the following factors and notations:
\begin{itemize}
	\item{\bf CT image for reconstruction:} The 3D image $\mu(\x,z)$ at point $(\x,z)=(x_1,x_2,z)\in \R^3$ (or tissue density) is reconstructed from measured projection data $\text{P}$.
	\item{\bf X-ray beam for CT:} A cone-shaped X-ray beam is passed through a patient positioned between an X-ray source and flat-panel detector. This beam is transmitted in different directions by rotating the gantry that houses the X-ray source and detector.
\item{\bf Radiation exposure:} CBCT devices emit a radiation dose in the range of 36.9-50.3 microsieverts ($\mu$Sv). Collimation, which limits the cross-sectional area of the X-ray beam to the area of the image receptor, reduces radiation exposure.
	\item{\bf CT projection data:} CBCT projection data $\text{P}(\varphi, u, v)$ are acquired using a planar detector after emitting the X-ray beams in several cone-beam directions, where $\Theta_{\varphi} =(\cos\varphi,\sin\varphi),~\varphi\in [0,2\pi)$ is the projection angle, and $(u,v)$ is the scaled planar detector position with scaling factor $\frac{R}{r}$. Here, $r$ is the source-to-rotation axis distance and $R$ is the source-to-detector plane distance, as illustrated in Fig. \ref{CBCTGeometry}. We define the cone-beam transform $\msC\mu(\varphi, u, v)$ as
\begin{equation}\label{CB-transform}
		\msC\mu(\varphi, u, v)=\int_0^{\infty}\mu(r(\Theta_{\varphi}^\perp,0) + t \omega_{u,v})dt,
\end{equation}
where $\Theta_{\varphi}^{\perp}=(\sin\varphi,-\cos\varphi)$, $(r\Theta_{\varphi}^\perp,0)$ describes how the source position moves around a source circle of radius $r$, and $\omega_{u,v}$ denotes the cone-beam direction starting at $(r\Theta_{\varphi}^\perp,0)$ and ending at detector position $(u,v)$. In addition, $\msC\mu(\varphi, u, v)$ is the line integral of $\mu$ along the cone-beam line associated with $(\varphi, u, v)$. The relationship between data location $(\varphi, u, v)$ and image location $(\x,z)$ is as follows:
\begin{align}\label{CBCT-0}
	u=r\f{\x\cdot\Theta_{\varphi}}{U_{\varphi, \x}}~~\text{and}~~v=z\f{r}{U_{\varphi, \x}} \quad \mbox{where  } U_{\varphi, \x} = r + \x \cdot \Theta_{\varphi}^\perp.
\end{align}
The measured data $\text{P}$  can be expressed as
\begin{equation}\label{Beerslaw-CBCT}
	\text{P}(\varphi, u, v) = -\mbox{ln} \int \eta(E)\exp(-\msC\mu_\Et(\varphi, u, v) ) dE + \textbf{noise},
\end{equation}
where  $\eta(E)$ represents the fractional energy at photon energy $E$ in the spectrum of the X-ray source with $\int_\R \eta(E) dE=1$, as shown in Fig. \ref{CBCTForwardRecon}.

\item{\bf Inverse Problem:} The goal of dental CBCT is to recover a proper tomography image $\mu$ from $\text{P}$ using the nonlinear relation \eqref{Beerslaw-CBCT}. Our aim is to reconstruct $\mu$ such that $\mu = \mu_{E_0}$ for a fixed energy level $E_0$.
\end{itemize}

Let us begin by reconstructing $\mu(\textbf{x},0)$ from the projection data  $\text{P}$ at level $v=0$. For ease of explanation, we transform $\text{P}(\varphi, u, 0)$ into the corresponding parallel beam data $\text{P}^{\sharp}(\theta,s)$ as follows:
\begin{equation}\label{Para-Fan-rel2}
	\text{P}^{\sharp}(\underbrace{\gamma_u + \varphi}_{\theta},\underbrace{r\sin\gamma_u}_{s}) = \text{P}(\varphi, u, 0),
\end{equation}
where $\gamma_u$ is the fan angle given by $u = r\tan\gamma$ and $\varphi+\f{\pi}{2}$ is the beam-source angle, as shown in Fig. \ref{CBCTGeometry}. The identity \eqref{Para-Fan-rel2} is derived from the following relation between the fan-beam line $\ell_{\varphi,u}^{\mbox{\tiny fan}} := \{\x\in \R^2~:~\x\cdot\Theta_{\varphi}=r\sin\gamma_u\}$ and parallel-beam line  $\ell_{\theta,s}^{\sharp}:= \{\x\in \R^2~:~ \x\cdot \Theta_{\theta}=s\}$ as follows:
\begin{equation}\label{ch3_Para-Fan-rel}
	\ell_{\theta,s}^{\sharp} = \ell_{\varphi,u}^{^{\mbox{\tiny fan}}} ~~\Longleftrightarrow~~s = r\sin\gamma_u, ~~ \theta = \gamma_u + \varphi.
\end{equation}
Under the ideal assumption of $\eta(E) = \delta(E-E_0)$ (a monochromatic beam) and no noise, we have
\begin{equation} \label{ideal_parallel}
	\text{P}^{\sharp} =  \mR\mu_{E_0},
\end{equation}
where the Radon transform $\mR\mu_{E_0}$ is defined by
\begin{equation}
	\mR\mu_{E_0}(\theta,s) = \int_{\mathbb{R}^2} \mu_{E_0} (\textbf{x},0) \delta(\textbf{x} \cdot \Theta_{\theta} - s) d\textbf{x}.
\end{equation}
Under the above ideal assumption, we have the FBP algorithm, given by
\begin{equation} \label{FBP}
	\mu_{E_0}(\textbf{x},0) = \mR^{-1} \text{P}^{\sharp}(\textbf{x}) = \dfrac{1}{8\pi^2} \int_{0}^{2\pi} \int_{\mathbb{R}} |\xi|\mathscr F_{s}[\text{P}^{\sharp}(\theta,\cdot)](\xi) \exp(i \xi \textbf{x}\cdot \Theta_{\theta}) d\xi d\theta,
\end{equation}
where $\mathscr F_{s}$ is the 1D Fourier transform with respect to variable $s$. However, in practice, because of the polychromatic nature of X-ray beam, $\text{P}^{\sharp}$ in \eqref{ch3_Para-Fan-rel} may not lie in the range space of the Radon transform, violating \eqref{ideal_parallel}. Hence, in the presence of a metallic material whose attenuation coefficient varies significantly with $E$, the FBP reconstruction $\mR^{-1} \text{P}^{\sharp}$ may produce severe artifacts, as shown in Fig. \ref{CBCTForwardRecon}.

The standard CBCT reconstruction algorithm is the FDK method, which was developed by Feldkamp, Davis, and Kress \cite{Feldkamp1984}, and can be regarded as an empirical 3D extension of the standard 2D parallel-beam FBP in \eqref{FBP}. The FDK algorithm  consists of filtering and weighted back-projection. The FDK algorithm computes the attenuation value $\mu(\x,z)$ at an image location $(\x,z)$ by suitably integrating  $\text{P}(\varphi,u,v)$ over all data positions $(\varphi, u, v)$ that relate to the beam lines passing through $(\x,z)$. For CBCT with a circular trajectory, as shown in Fig. \ref{CBCTGeometry}, the FDK algorithm for reconstructing $\mu$ can be expressed as
\begin{equation}\label{conbeam-recon2}
\FDK[\text{P}](\x,z)  =\f{1}{4\pi}\int_0^{2\pi} \f{r^2}{U_{\varphi,\x}^2} \int \text{P}(\varphi,u,\frac{zr}{U_{\varphi, \x}}) \dfrac{r ~ \hbar(r\frac{\x \cdot \Theta_\varphi}{U_{\varphi,\x}}-u)}{\sqrt{r^2+u^2+(\frac{zr}{U_{\varphi, \x}})^2}}dud\varphi,
\end{equation}
where $\hbar(u)$ is a 1D ramp filter given by $\hbar(u)=\displaystyle \dfrac{1}{2\pi} \int_{\mathbb{R}} |\xi|e^{i \xi u} d\xi.$ For details of the CBCT reconstruction, we refer the reader to \cite{Grangeat1991,Katsevich2001}.

The ill-posedness of the inverse problem (i.e., recover $\mu$ from  $\text{P}$) is related to the following commercial dental CBCT specifications: circular cone-beam scan, a scan time of 5-24s, FOV truncation, offset detector, low X-ray dose, and a cost of less than 100 thousand USD. Meanwhile, MDCT uses a helical fan-beam scan and has a scan time that is less than 1s, no FOV truncation, no offset detector, a high X-ray dose, and a cost of over 1 million USD.

The main parameters influencing the radiation dose in a given CBCT device are the tube current, tube voltage, and collimation. To reduce the radiation exposure in CBCT, it is recommended to use as small field of view (FOV) as possible, the lowest tube current setting, and the shortest exposure time. CBCT devices generally emit 36.9-50.3 $\mu$Sv of radiation dose. FOV truncation is where the size of the FOV is determined by the detector size and shape, beam projection geometry, and beam collimation function, as illustrated in Fig. \ref{cone-beam}. Because a significant portion of the manufacturing cost of a CBCT device is the cost of the  X-ray detector, the smallest detector is used to obtain the desired image.

\begin{remark} Many researchers misunderstand that current dental CBCTs have a higher spatial resolution than MDCT. This misconception is based on the fact that recent dental CBCTs use flat detectors with a pixel size of $0.1mm\times 0.1mm$, whereas current MDCTs have spatial resolution of greater than 0.2mm. We must distinguish between ``nominal spatial resolution" in non-living objects and ``actual spatial resolution" in living patients. The actual spatial resolution of real-world dental CBCTs is negatively affected by multiple factors including patient motion owing to the long acquisition time, X-ray scattering, various sources of noise and artifacts caused by low X-ray dose and FOV truncation, cone-beam reconstruction errors caused by violation of the Tuy condition \cite{Tuy1983}, and others. We note that a smaller detector size tends to lower the signal-to-noise ratio. Br\"ullmann \etal\cite{Brullmann2015} stated that ``Voxel size is also commonly mistaken as spatial resolution. Technically, the spatial resolution of CBCT devices is related to the physical pixel size of the sensor, the grey-level resolution, the reconstruction technique applied and some other factors. Many additional parameters affect the image quality and exposure doses, such as exposure parameters, tube voltage, tube current, exposure time and rotation arc."
\end{remark}

\subsection{Ill-posedness issues}
\subsubsection{Underdetermined problem: Truncated FOV}
Most dental CBCTs use a FOV that is smaller than that of a patient's body, resulting in significant truncation of the projection data. As shown in Fig. \ref{cone-beam} (b), the effective FOV of dental CBCT does not cover the entire region of the object to be scanned and offset detector geometry is used. The dental CBCT sinogram $\text{P}$ can be expressed as
\begin{equation} \label{pfull}
	\text{P} = \mathcal S_{\mbox{\scriptsize ub}}( \text{P}_{\mbox{\scriptsize full}}),
\end{equation}
where $\text{P}_{\mbox{\scriptsize full}}$ is the corresponding sinogram acquired with a non-offset and wide-detector CBCT, providing the entire information of a sinogram, and $\mathcal S_{\mbox{\scriptsize ub}}$ is a subsampling operator determined by the size and offset configuration of the detector. This missing information in $\text{P}$ along the $u$-axis leads to the need to solve an undersampled problem (the so-called interior tomography problem \cite{Natterer1986}). This interior problem is known to have no unique solution in general \cite{Natterer1986}.

To explain the interior tomography, let us consider the simple 2D parallel CT model data $\text{P}^{\sharp}$ in \eqref{Para-Fan-rel2}, which corresponds to CBCT data $\text{P}$ at detector position $v=0$. We consider the problem of recovering $\mu(\x)=\mu(\x,0)$ in the ROI area $\Om_{\mbox{\tiny ROI}}=\big\{\x\in\R^2:|\x|\leq d_{\mbox{\tiny ROI}} \big\}$ from the projection data $\text{P}^\sharp$ given in the truncated area $\Pi_{\mbox{\tiny ROI}} = [-\pi,\pi)\times[-l_{\mbox{\tiny ROI}},l_{\mbox{\tiny ROI}}]$. The following directional Hilbert transform is used to address the interior tomography problem \cite{Noo2004,Wang2013}:
\begin{align}
	\mH_{\theta_0}\mu(\x)&=\f{1}{\pi}~p.v.\int_{-\infty}^{\infty}\f{\mu(\x-t\Theta_{\theta_0})}{t}dt.\label{directional_Hilbert}
\end{align}
where $\theta_0$ is a fixed angle and $p.v.$ denotes the Cauchy principal value. Note that if $\mR \mu$ is known for every $(\theta,s)\in(-\pi,\pi]\times\R$, then $\mu(\x)$ can be recovered using the following identity \cite{Noo2004,Wang2013}
\begin{align}\label{BFP}
	\mu(\x)=\f{1}{2\pi}\mH_{\bth_0}\mR_{\bth_0}^*\left[\f{\p}{\p s}\mR \mu\right](\x),
\end{align}
where $\mR^*_{\theta_0}$ denotes the backprojection operator depending on $\bth_0$, and is defined by
\begin{align}
	\mR_{\theta_0}^*g(\x)&=\int_{|\theta-\theta_0|<\frac{\pi}{2}} g(\theta,\x\cdot \Theta_{\theta})d\theta \label{backprojection}.
\end{align}
Under the ideal monochromatic assumption \eqref{ideal_parallel}, $\mR \mu$ can be replaced by $\text{P}^\sharp_{\mbox{\tiny full}}$, which is a full-sampled projection data corresponding to $\text{P}^\sharp$. Then, \eqref{BFP} can be rewritten as
\begin{align}
	\mu(\x)=\f{1}{2\pi}\mH_{\bth_0}\mR_{\bth_0}^*\left[\f{\p}{\p s}\text{P}^\sharp_{\mbox{\tiny full}}\right](\x).
\end{align}
According to \cite{Natterer1986}, there exists a non-zero $\widetilde{\mu}$ in $\Om_{\mbox{\tiny ROI}}$ such that $\mR \widetilde{\mu} = 0$ in $\Pi_{\mbox{\tiny ROI}}$. To be precise, we can find a function $h \in C^\infty(\mathbb{R})$ satisfying the Helgason-Ludwig condition (see Theorem 4.2 in \cite{Natterer1986}) such that $h(s) = 0$ for $|s|<l_{\mbox{\tiny ROI}}$. Then, $\widetilde{\mu}$ can be reconstructed by
\begin{align} \label{mutilde}
	\widetilde{\mu}(\x)=\f{1}{2\pi}\mH_{\bth_0}\mR_{\bth_0}^*\left[\f{\p}{\p s}\widetilde{\text{P}}^\sharp_{\mbox{\tiny full}}\right](\x),
\end{align}
where $\widetilde{\mbox{\text{P}}}^\sharp_{\mbox{\tiny full}}(\theta,s)=h(s)$. This leads to the non-uniqueness of interior tomography, because
\begin{equation}
	\mR (\mu + \widetilde{\mu} ) = \mR \mu = \mbox{\text{P}}^{\sharp} ~~ \mbox{ in } \Pi_{\mbox{\tiny ROI}}.
\end{equation}
Note that, according to \eqref{mutilde}, $\widetilde{\mu}$ is analytic along the line segment $\Omega_{\mbox{\tiny ROI}} \cap \big\{\x+t\Theta_{\theta_0}:t\in\R\big\}$ because $\widetilde{\mbox{\text{P}}}^\sharp = 0$ in $\Pi_{\mbox{\tiny ROI}}$. Indeed, the reconstructed image $\mu$ in $\Omega_{\mbox{\tiny ROI}}$ is unique up to directional analytic images in $\Omega_{\mbox{\tiny ROI}}$ \cite{Noo2004,Defrise2006,Wang2013}.

\subsubsection{Nonlinear problem: X-ray energy-dependent attenuation}
The projection data $\text{P}$ in (\ref{Beerslaw-CBCT}) are nonlinear with respect to $\mu$ because the value of $\mu_\Et$ varies with $E$. The nonlinearity of $\text{P}$ depends on the geometries of the objects (e.g., tissues, bones, and metallic implants) and the energy spectrum $\eta$. Fig. \ref{CBCTForwardRecon} shows the energy spectrum of the X-ray beam and the attenuation coefficients of iron, bone, and tissue from the experimental results given in \cite{Hubbell1995}. For example, $\mu_{E}(\mbox{material})$ is:
\begin{itemize}
	\item $\mu_{30KeV}(\mbox{soft tissue})\approx 0.40 (cm^{-1}), ~\mu_{80KeV}(\mbox{soft tissue})\approx 0.19 (cm^{-1})$
	\item $\mu_{30KeV}(\mbox{bone})\approx 2.56 (cm^{-1}), ~\mu_{80KeV}(\mbox{bone})\approx 0.43 (cm^{-1})$
	\item $\mu_{30KeV}(\mbox{Iron})\approx 64.38 (cm^{-1}), ~\mu_{80KeV}(\mbox{Iron})\approx 4.69 (cm^{-1})$
\end{itemize}
The value $\mu$ of a soft tissue varies little with $E$, whereas the $\mu$ of a metal object varies greatly. Hence, in the presence of high attenuation materials (i.e., high $\frac{\p}{\p E}\mu_\Et$) such as metals in the FOV of a clinical CBCT, the FDK algorithm \eqref{conbeam-recon2} produces serious streaking artifacts, which are mainly caused by the beam hardening factor (the lower-energy photons tend to be absorbed more rapidly than higher energy photons) \cite{DeMan2001,Elbakri2002,OSullivan2007,Lee2019}. Recently, major CT companies have been developing photon-counting CTs that have the potential to overcome the above mentioned nonlinear issues of current CTs \cite{Danielsson2021,Flohr2020,Leng2019,Shikhaliev2005,Willemink2018}. However, it seems difficult to use a photon-counting detector for dental CBCT for the time being.

For ease of explanation and simplicity, we limit our discussion to the parallel-beam model data $\text{P}^{\sharp}$ in {\ref{Para-Fan-rel2}), which correspond to the CBCT data $\text{P}$ at detector position $v=0$.
An unavoidable mismatch exists between the nonlinear projection data $\text{P}^{\sharp}$ and the linear mathematical model based on the ideal monochromatic assumption (\ref{ideal_parallel}). Fig. \ref{metalartifact} shows the violation of the linear assumption \eqref{ideal_parallel}. The mismatch between the data $\text{P}^{\sharp}$ and the range space of $\mR$ can be explained by the polychromatic nature of the X-ray beams \cite{Natterer1986}.

For a rigorous description of the geometry of beam hardening, consider a simple cross-sectional body (corresponding to $z=0$) composed of soft tissue and metal.  Park \etal \cite{Park2015, Park2017} found
the nonlinear beam hardening factor of projection data $\text{P}^{\sharp}$ induced by metal. The reconstructed artifact image due to the metal-occupying region $D$ is given by
\begin{equation}\label{fma}
	\phi_{D,\lambda}(\x)=-\f{1}{8\pi^2}\int_{-\pi}^{\pi}\int_{-\infty}^\infty  |\om|\msF\left[\ln\left(\f{\sinh\left( \lambda\mR \chi_D(\theta,\cdot)\right)}{\lambda\mR \chi_D(\theta,\cdot)}\right)\right](\om)e^{i\omega \x\cdot\Theta_{\theta}}d\om d\theta,
\end{equation}
where $D$ is the metal region, $\lambda$ is a constant that depends on the energy spectrum of the X-ray beam and absorption property of the subject, and $\chi$ is a characteristic function.
The projection data $\text{P}^{\sharp}$ can be decomposed into data consistent and inconsistent parts according to the metal region $D$ as follows:
\begin{equation}
\text{P}^{\sharp}(\theta,s)=\underbrace{\mR\mu_{E_0}(\theta,s)}_{\mbox{target}}+\underbrace{\ln\left(\f{\sinh\left( \lambda\mR \chi_D(\theta,s)\right)}{\lambda\mR \chi_D(\theta,s)}\right)}_{\mbox{model mismatch}}.
\end{equation}
The beam hardening corrector $\phi_{D,\lambda}$ enables us to handle the serious mismatch between projection data $\text{P}^{\sharp}$ and mathematical model $\mR\mu_{E_0}$ in the presence of metallic objects. The idea of the geometric corrector is the selective extraction of metal-induced streaking and shadow artifacts from uncorrupted CT images without affecting intact anatomical images.

The first mathematical analysis to characterize the structure of metal streaking artifacts was presented in \cite{Park2017}. On the basis of this mathematical analysis, the authors first found the mathematical formula \eqref{fma} for beam-hardening metal artifacts, which was experimentally validated using industrial CT \cite{Park2015}. Metal artifacts are viewed as a singularity propagation in the image, which is closely related to the interrelation between the structure of data $\text{P}^\sharp$ and $\mu = \mR^{-1} \mbox{\text{P}}^\sharp$. This can be interpreted effectively using the Fourier integral operator and wavefront set \cite{Duistermaat1972,Hormander1971,Hormander1983,Petersen1983,Tr`eves1980}.

The following theorem characterizes metal artifacts in terms of their geometry:

\begin{theorem} (Park \etal. \cite{Park2017}) \label{th32} Let $D_1,D_2,\ldots,D_J$ be strictly convex and disjoint  bounded domains in $\R^2$ with connected boundaries. Let $D=\cup_{j=1}^J D_j$ be the metal region.  Given $\text{P}$, assume that $\mu$ is represented as
\begin{align}\label{fma3-1}
\mu(\x)=\mu_{E_0}(\x)+\Upsilon_{\text{P}}(\x),
\end{align}
where
\begin{align}\label{fma3-2}
\Upsilon_{\text{P}}=\f{1}{4\pi}\mR^*\mI^{{\tiny\mbox{$-1$}}}\left[\sum_{k=1}^N\f{(-1)^{k} }{k}\left[\sum_{n=1}^{N}\f{(\alpha\eps)^{2n}}{(2n+1)!}(\mR \chi_D)^{2n}\right]^k\right].
\end{align}
If a straight line $\ell_{\theta,s}=\big\{\x=s\Theta_{\theta}+t\Theta_{\theta}^\perp: t\in \R\big\}$ satisfies
\begin{align}\label{artifact0}
	\Sigma_{\x}(\mu)\neq \emptyset \q\mbox{for all }~ \x\in \ell_{\theta,s}\setminus \ssu(\mu_{E_0}),
\end{align}
then $(\theta,s)$ satisfies
\begin{align}\label{streaking3}
\dim\left(\mathrm{Span} [ \Sigma_{(\theta,s)}(\mR\chi_D)]\right)=2,
\end{align}
where $\dim(\mathrm{Span} [A])$ is the dimension of the span of the set $A$, $\ssu(u)$ denote the singular support of $u$, and $\Sigma_{\x}(u)$ is a closed conic subset in $\R^2\setminus\{{\bf 0}\}$ \cite{Hormander1983,Tr`eves1980}.
\end{theorem}

Fig. \ref{metalartifact} provides a visual explanation of Theorem \ref{th32}. Metal streaking artifacts are produced only when the wavefront set of the Radon transform of $\sum_{j=1}^N\chi_{D_j}$  does not contain the wavefront set of the square of the Radon transform. Metal streaking artifacts can appear when there exist distinct $\y$, $\z\in\cup_{j=1}^N \p {D_j}$ such that the straight line $l_{\theta,s}$ is tangent to the boundaries $\cup_{j=1}^N \p {D_j}$ at $\y$ and $\z$ simultaneously.

\subsubsection{Cone beam artifacts and violation of Tuy's data complete condition}
\begin{figure}[h]
	\centering
	\includegraphics[width=1\textwidth]{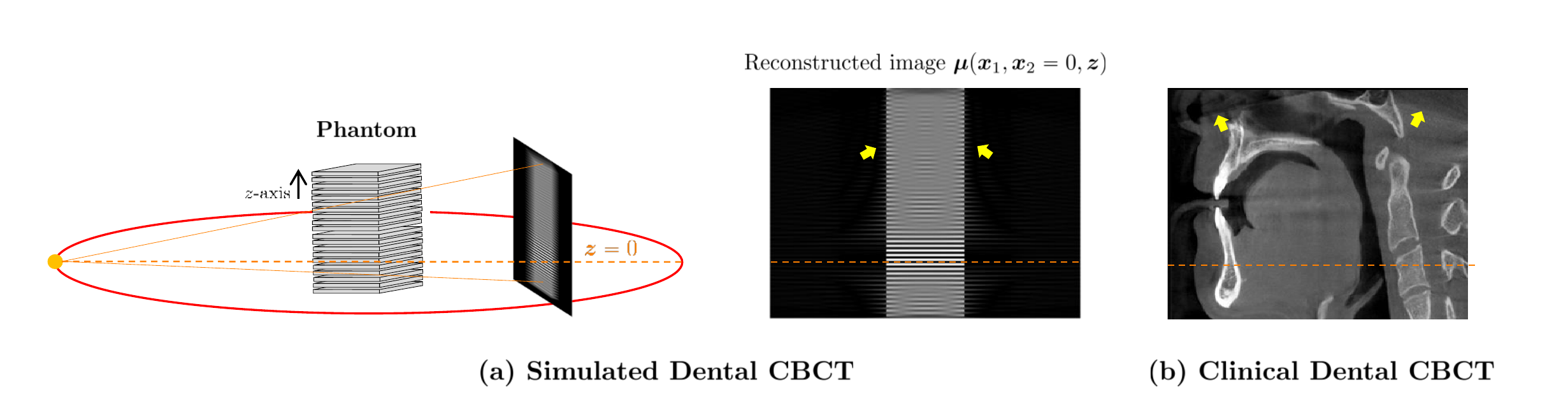}
	\caption{Cone-beam artifacts due to the violation of Tuy's condition. It can be seen in the yellow arrows that the cone-beam artifacts increase as $\boldsymbol z$ moves away from $\boldsymbol z = 0$ (the orange dashed line).}
	\label{ConeBeamArtifacts}
\end{figure}

Cone-beam artifacts are caused by the violation of the data sufficiency condition formulated by Tuy in \cite{Tuy1983}. According to Tuy's condition, the accurate reconstruction of a plane requires that the plane contains the X-ray's focal spot point. In dental CBCT, the reconstructed image $\FDK[\text{P}]$ contains these cone-beam artifacts. In circular CBCT, as shown in Fig. \ref{ConeBeamArtifacts}, the projection data $\text{P}$ (even with the monochromatic assumption) is insufficient for accurate analytic reconstruction, because it is not true that every plane passing through any location in the region of interest (ROI) intersects the source trajectory at least once. Most dental CBCT machines violate Tuy's condition, whereas most MDCTs are designed to meet Tuy's condition.

\section{Methods for solving the ill-posed problems in dental CBCT}
To solve the ill-posed problem in dental CBCT, the goal is to find a nonlinear reconstruction map $f: \text{P}\mapsto \mu_*$  that maps from the noisy under-sampled data $ \text{P}$ to the corresponding CBCT image $\mu_*$ that we want to reconstruct. In addition, we should take into account photon starvation, which is very common in dental CBCT,  especially when the patient has many implants.
To solve this ill-posed problems, a regularized least squares method of the following form can be used:
\begin{equation}\label{leastSquare1}
	\mu_*=\underset{\mu}{\mbox{argmin}} \f 12  \|\text{P} - \msC\mu\|_{\ell_2}^2 +\lambda \Gamma (\mu),
\end{equation}
where $\Gamma (\mu)$ is a regularization term constraining prior knowledge of artifact-free and noise-free CBCT images,  $\msC$ is the cone-beam transform in \eqref{CB-transform}, $\|\cdot\|_{\ell_2}$ denotes the standard Euclidean norm, and $\lambda$ is the regularization parameter controlling the trade-off between the fidelity term $ \|\text{P} - \msC\mu\|_{\ell_2}^2 $ and regularity. This regularization should help suppress artifacts in the reconstructed image. Using the FDK algorithm $\FDK[\text{P}]$ defined in (\ref{conbeam-recon2}), the least-squares problem \eqref{leastSquare1} can be expressed as
\begin{equation} \label{leastSquare2}
	\mu_*=\underset{\mu}{\mbox{argmin}} \f 12  \|\FDK[\text{P}] - \mu\|_{\ell_2}^2 +\lambda \Gamma (\mu).
\end{equation}

The linear model $\msC\mu$ or $\FDK[\text{P}]$ is based on the unrealistic assumption that $\eta(E) = \delta(E-E_0)$ (monochromatic beams) and a serious discrepancy in the fidelity term can occur in the presence of metallic objects. As a good approximation of the polychromatic beam \eqref{Beerslaw-CBCT}, a more realistic model would be to replace $\mathcal C \mu$ with the following cone beam projection $\widehat{\mathcal C} \mu$ based on Alvarez's assumption \cite{Alvarez1976}:
\begin{equation}\label{realCone}
	\widehat{\mathcal C} \mu = -\mbox{ln} \int \eta(E)\exp(\underbrace{-p(E)\psi_{p}(\mu)-q(E)\psi_{q}(\mu)}_{-\mu_{\Et}} ) dE,
\end{equation}
where $\psi_{p}(\mu)$ is the spatially-dependent photoelectric component, $\psi_{q}(\mu)$ is the spatially-dependent Compton scattering component, $p(E)$ approximates the energy dependence of the photoelectric interaction, $q(E)$ is the Klein-Nishina function \cite{Klein1929,Stonestrom1981}, and $\mu$ represents the monochromatic linear attenuation coefficient at $E = 70 keV$ \cite{DeMan2001}. Here, $\psi_{p}$ and $\psi_{q}$ are assumed to be known functions of $\mu$.

The key to finding the nonlinear reconstruction map $f$ is to determine a good regularization model $\Gamma (\mu)$ that penalizes $\mu$ based on how far it is from the prior distribution of the expected images.

\subsection{Classical methods for MAR}
Over the past four decades, numerous MAR methods have been developed based on the MDCT model with full FOV projection data. These include statistical iterative correction \cite{DeMan2001,Elbakri2002,Menvielle2006,OSullivan2007,Williamson2002}, sinogram inpainting-based correction \cite{Abdoli2010,Bazalova2007,Kalender1987,Meyer2010,Park2013}, and dual-energy approaches \cite{Alvarez1976,Lehmann1981,Yu2012}. Most commercial MAR algorithms can be considered to be hybrid methods that combine iterative methods and sinogram correction approaches. These include SEMAR (Toshiba Medical Systems) \cite{Zhang2017}, O-MAR (Philips Healthcare) \cite{Omar2012}, iMAR (Siemens Healthcare) \cite{Kachelriess2016}, and Smart MAR (GE Healthcare) \cite{Gemar2013}. Sinogram correction approaches use various inpainting techniques such as interpolation, reprojection, and normalization to recover unreliable background data because of the presence of metallic objects.

We briefly explain the sinogram correction method using TV inpainting \cite{Duan2008}. Given a reconstructed image $\mu$, we segment a metal region $D$ by using a suitable threshold, and then obtain the corresponding metal trace $T$ given by $T = \mbox{supp}{\{\mathcal C \chi_{D}\}}$. Sinogram correction $\mbox{\text{P}}_{\mbox{\footnotesize cor}}$ can be obtained by minimizing the following objective function:
\begin{equation}
	\mathcal L_{M} (\mbox{\text{P}}_{\mbox{\footnotesize cor}}) := \dfrac{1}{2}\|(\mbox{\text{P}}_{\mbox{\footnotesize cor}} - \mbox{\text{P}})\odot \chi_{T^c} \|^2_{\ell^2}  + \lambda \|\nabla \mbox{\text{P}}_{\mbox{\footnotesize cor}}\|_{\ell^1},
\end{equation}
where $\odot$ is Hadamard's product, $\chi_{T^c}$ is the binary mask of the sinogram area excluding the metal trace $T$, and  $\|\cdot \|_{\ell^1}$ is the $L^1$-norm. Unfortunately, these methods can create new artifacts, and tend to corrupt the morphological information of the region around the metallic object in the reconstructed image. Hence, these approaches  may not be suitable for dental CBCT, where individual tooth details are more important than the overall structure.

Normalized MAR (NMAR) \cite{Meyer2010} is widely used to handle streaking artifacts caused by the non-smooth transition between the original and interpolated data. The main idea of NMAR is to convert the sinogram to a nearly flat sinogram and then interpolate it to obtain a smooth transition. However, the NMAR method does not recover the tooth structure near a metallic object, because its performance depends on the accuracy of the object segmentation. A fundamental limitation of this method is that the segmentation, which is based on thresholds, is not perfectly accurate.

The other solution, dual-energy CT, is not appropriate for low-dose dental CBCT, because the use of an additional scan with a higher energy increases radiation exposure to patients. Similarly, photon-counting detector that provides monochromatic data by differentiating individual photon energies  is not suitable for low-cost dental CBCT.

\subsection{Iterative reconstruction}
In the variational approach \eref{leastSquare1} to solving the inverse problems of X-ray CT, numerous regularizations reflecting the spatial correlation among neighboring pixels, such as TV \cite{Dong2014,Xu2011}, fractional-order TV \cite{Zhang2016}, nonlocal TV \cite{Liu2016}, Markov random field theory \cite{Sauer1993,Li2004,Wang2006}, and nonlocal means \cite{Chen2009,Kim2017,Ma2012,Zhang2017}, have been used on the target image. However, it is challenging to design a regularization method that conveys the characteristics of target image $\mu^*$. For example, in compressed sensing (CS) \cite{Candes2005,Donoho2006}, $\Gamma(\mu)$ can be $L^1$-norm that increases sparsity in a given basis. These CS-based regularizations produce an over-smoothing effect that causes the loss of fine tooth detail. The internal parameters of the CS-based regularizations also have a significant impact on the reduction of artifacts.

To make this explanation concrete, we denote the projection operator by a matrix $\A=[a_{ij}]$, where $a_{ij}$ is the effective intersection length of the $i-$th projection line with the $j-$th voxel. We use the bold face letters $\mmu\in\R^N$ and $\mathbf{P}\in\R^M$, to denote the corresponding discrete versions of $\mu$ and $\text{P}$, respectively. The least squares problem corresponding to \eref{leastSquare1} is to minimize the following:
\begin{equation}\label{leastSquare3}
	\mathcal L(\mmu)=\f{1}{2}\| \A \mmu-\mathbf{P}\|_{\mathbf{H}}^2 + \lambda \Gamma(\mmu),
\end{equation}
where $\|\mmu\|_{\mathbf{H}}=\mmu^T\mathbf{H}\mmu$ and $\mathbf{H}=\I-\mathbf{T}$. Here, $\I$ is an identity matrix and $\mathbf{T}=diag(t_i)$ is a diagonal matrix whose diagonal elements $t_i$ are one on the metal trace and zero otherwise. In the presence of metallic objects, the fidelity domain in \eref{leastSquare3} is restricted to the outside metal region to avoid serious discrepancy in the fidelity.

The proximal-gradient method provides the following iterative procedure: For each $k=0,1,2,\ldots, K$,
\begin{equation} \label{mu_1}
\mmu^{(k+1/2)} = \mmu^{(k)} - \gamma \nabla \Gamma(\mmu^{(k)})  ~~\text{and}~~ \mmu^{(k+1)} = g(\mmu^{(k+1/2)},\mathbf{P}),
\end{equation}
where $\gamma$ ia the step size at iteration $k$ and $g(\mmu, \mathbf{P})$ is the function given by
\begin{align}\label{prox-grad}
  g(\mmu, \mathbf{P}) = \arg\min_{\tilde\mmu} \mathcal L(\tilde\mmu;\mmu,\mathbf{P}) =  \|\A\tilde\mmu-\mathbf{P}\|_{\mathbf{H}}^2 + \f{\lambda}{\gamma}\|\tilde\mmu-\boldsymbol{\mu}\|_2^2.
\end{align}

The minimization problem of \eref{prox-grad} can be solved using the following separable paraboloid surrogate \cite{Elbakri2002}:  \begin{equation}
Q(\tilde{\mmu};\tilde{\mmu}^{(l)},\mmu,\mathbf{P})=\sum_{i=1}^{M}\sum_{j=1}^N \beta_{ij}h_i\left(\f{a_{ij}}{\beta_{ij}}(\tilde{\mu}_j-\tilde{\mu}^{(l)}_j)+\sum_{j=1}^N a_{ij}\tilde{\mu}_j^{(l)}-P_i\right)^2+\f{\lambda}{\gamma}\sum_{j=1}^{N}(\tilde{\mu}_j-\mu_j)^2,
\end{equation}
where $\beta_{ij} = a_{ij}\slash\sum_{j=1}^N a_{ij}$.
Then, $Q(\tilde{\mmu};\tilde{\mmu}^{(l)},\mmu,\mathbf{P})$ satisfies
\begin{align}
  \mL(\tilde{\mmu};\mmu,\mathbf{P})\leq Q(\tilde{\mmu};\tilde{\mmu}^{(l)},\mmu,\mathbf{P}),~
  \mL(\tilde{\mmu}^{(l)};\mmu,\mathbf{P})= Q(\tilde{\mmu};\tilde{\mmu}^{(l)},\mmu,\mathbf{P}).
\end{align}
We minimize the paraboloid $Q(\z;\z^{(l)},,\mmu,\mathbf{P})$ instead of $\mL(\z;\mmu,\mathbf{P})$. Since $Q(\z;\z^{(l)},\mmu,\mathbf{P})$ is a separable paraboloid, it can be explicitly optimized as follows: For each pixel $j=1,2,\ldots, N$,
\begin{equation}\label{EQ}
  \tilde{\mu}_j^{(l+1)} = \tilde{\mu}_j^{(l)} - \f{\displaystyle \sum_{i=1}^M a_{ij}h_i\left(\sum_{j=1}^N a_{ij}\tilde{\mu}_j^{(l)}-P_i\right) + \f{\lambda}{\gamma}(\tilde{\mu}_j^{(l)}-\mu_j)}{\displaystyle \sum_{i=1}^M \f{a_{ij}^2 }{\beta_{ij}} + \f{\lambda}{\gamma}}.
\end{equation}
The \eqref{EQ} can be simply calculated as the following matrix-vector multiplication:
\begin{align}\label{z-update}
  \tilde{\mmu}^{(l+1)} = \tilde{\mmu}^{(l)} -\f{\A^T\mathbf{H}(\A\tilde{\mmu}^{(l)}-\mathbf{P})+\f{\lambda}{\gamma} (\tilde{\mmu}^{(l)}-\mmu)}{\A^T\mathbf{H}\A\boldsymbol{1} + \f{\lambda}{\gamma} \boldsymbol{1}},
\end{align}
where $\A^T$ is the transpose of $\A$ (i.e., the back projection) and $\boldsymbol{1}$ is a vector of ones.

\subsection{Data-driven regularization instead of handcrafted regularization}
In the iterative reconstruction in \eref{mu_1}, the impact of regularization (i.e., $-\gamma \nabla \Gamma (\mmu^{(k)})$) plays a key role in the success of reconstructions. Here, two questions arise: (i) which regularization is the most appropriate, and (ii) whether it is appropriate to use the same regularization in each iteration. Because artifacts and noise characteristics differ at each iteration step, it is desirable to use a different artifact corrector for each iteration step. Hand-craft regularization priors, such as TV, seem to have limited performance in handling artifacts that have nonlinear structures depending on metal geometry.

Recently, data-driven regularization was used to compute $\mmu^{(k+1/2)}$ in \eref{mu_1} \cite{Park2022}, where the regularization function was different for each iteration step. This is a supervised learning method to find an artifact extractor $f^{(k)}: \mmu^{(k)}\rightarrow \boldsymbol{\zeta}^{(k)}$, where $\boldsymbol{\zeta}^{(k)}$ indicates artifacts included in the reconstructed image $ \mmu^{(k)}$ at the $k-$th iteration step.  Here, $\gamma\nabla\Gamma(\mmu^{(k)})$ in \eref{mu_1} is replaced by  $f^{(k)}(\mmu^{(k)})$. The artifact extractor $f^{(k)}(\cdot; \W^{(k)})$ is a trainable neural network that depends on parameters $W^{(k)}$. See \cite{Park2022} for details of the network architecture of $f^{(k)}(\cdot; \W^{(k)})$.  Assuming that paired data $\mS^{(k)} = \{(\mmu^{(k)}_{i},\boldsymbol{\zeta}^{(k)}_{i})~|~i=1,2,\ldots,L\}$ is available, the network $f^{(k)}(\cdot; \W^{(k)})$ is trained using the paired training dataset through the following framework:
\begin{align}\label{direct-m}
	\W^{(k)} = \arg\min_{\W}\f{1}{L}\sum_{i=1}^L\|f^{(k)}(\mmu^{(k)}_i ; \W)-\boldsymbol{\zeta}^{(k)}_{i}\|_2^2,
\end{align}
where  $\|\cdot\|_2$ denotes the standard Euclidean norm.

In practice, it is difficult to directly obtain a dataset consisting of artifact-free $\mmu^*$ paired with the corresponding artifact $\boldsymbol{\zeta}$. However, large amount of unpaired data are easy to obtain. Hence, we can combine real data and metal artifact simulation to generate paired data. First, we obtain artifact-free data from metal-free patients. Second, we perform individual tooth segmentation on the artifact-free data using the deep learning-based segmentation technique in \cite{Jang2021} and  choose several tooth positions in which virtual metal implants could be placed. Third, we generate a variety of sinogram data affected by the different geometries of metal implants using the accurate forward model in \eqref{Beerslaw-CBCT} \cite{Park2018,Zhang2018-1}. Finally, various data induced by simulated-metals are added to the metal-free data. This process provides a paired dataset, $\mS^{(0)} = \{(\mmu^{(0)}_{i},\boldsymbol{\zeta}^{(0)}_{i})~|~i=1,2,\ldots,L\}$.

This paired dataset $\mathcal{S}^{(0)}$ is used to train $f^{(0)}$ via \eref{direct-m}. Next, $\mathcal{S}^{(0)}$  and $f^{(0)}$ are used to generate $\mS^{(1)}=\{(\mmu_i^{(1)},\boldsymbol{\zeta}_i^{(1)})~|~i=1,2,\ldots,L\}$, where $\boldsymbol{\zeta}_i^{(1)}= \mmu_i^{(1)}-\mmu_i^*$ and
\begin{equation}
	\mmu_i^{(1)}=g(\mmu_i^{(0)}-f^{(0)}(\mmu_i^{(0)};\W^{(0)})~,~\mathbf{P}_i).
\end{equation}
This process continues iteratively to obtain $\mS^{(k)}, k=2,\ldots,K$.
Fig. \ref{fig:results_DL} shows the results of the data-driven regularization approach.

\section{Image priors}
Throughout this section, the notation $\z$ is used to represent an image $\z=\FDK[\mathbf{P}]$ reconstructed by the FDK algorithm \eref{conbeam-recon2}. Let  $p_{*}(\mmu)$ represent the probability distribution of artifact-free CBCT images. Our aim is to model $p_{*}(\mmu)$ such that when the $\mmu$ is closer to an artifact-free CBCT image, a higher $p_{*}(\mmu)$ is assigned.
Given  $\z=\FDK[\mathbf{P}]$, the least squares problem \eqref{leastSquare2} can be considered to be equivalent to the following maximum a posteriori (MAP) estimation \cite{Ulyanov2020,Veen2020}:
\begin{equation} \label{Prior1}
	\mmu^*=\underset{\mmu}{\mbox{argmin}} (- \log p( \z ~|~\mmu) - \log p_{*}(\mmu)),
\end{equation}
where $p_{*}(\mmu)=\exp(-\Gamma(\mmu))$ can be viewed as a prior and the conditional distribution $p(\z ~ |~\mmu)=\exp(\|\z- \mmu\|_2^2/2\lambda)$ is regarded as the data fidelity. The challenge is then how to assign the prior $p_{*}(\mmu)$.

Assume that any artifact-free dental CBCT image lies near or on an (unknown) manifold $\mathcal{M}_*$, whose Hausdorff dimension is much smaller than the dimension of the sample space (i.e., the number of voxels in the dental CBCT image). The role of $p_{*}(\mmu)$ is to apply a force that causes $\mmu$ to lie in or near manifold $\mathcal{M}_*$. For example, $p_*(\mmu)$ can be $p_*(\mmu)\propto \exp(-\mbox{dist}(\mmu,\mathcal{M}_*))$, where $\mbox{dist}(\mmu,\mathcal{M}_*)$ is a suitable distance between $\mmu$ and $\mathcal{M}_*$ from the perspective of dental radiologists. In CS approaches,  $\mathcal{M}_*$ is assumed to have a locally very sparse dimension, that is, $\mmu\in \mathcal{M}_*$ is sparse in a suitable basis. These CS approaches include TV, which imposes the sparsity of the image gradient using the $\ell^1$-convex relaxation $\Gamma(\mmu)=\|\nabla \mmu\|$. However, CS methods cannot selectively preserve small details, because $\Gamma(\mmu)$ penalizes uniformly based on a fixed sparsity in a given basis. These approaches lack the ability to contextually control the anatomical details, which tends to compromise the morphological information in a region around metallic objects while reducing noise and artifacts.

Deep learning approaches such as generative adversarial networks \cite{Goodfellow2014} have shown remarkable performance in regressing manifold  using training data. Here, training data are used to train a generator $G$ such that $\mathcal{M}_*\approx \{ \mmu=G(\z)~:~  \z \sim p_{\mbox{\tiny FDK}}\}$, where $p_{\mbox{\tiny FDK}}(\z)$ represents the probability distribution of the reconstructed images using the FDK algorithm \eqref{conbeam-recon2}. This generator $G: \z \mapsto \mmu \sim p_{*}$ can be viewed as an artifact correction function that minimizes the following loss model:
\begin{align}\label{model0}
	\mbox{dist}(p_G,p_*) + \lambda E_{\z\sim p_{\mbox{\tiny FDK}}} \left[ \| G(\z) - \z\|_2^2\right],
\end{align}
where $p_G$ is the distribution of the generated images $G(\z),~\z\sim p_{\mbox{\tiny FDK}}$ and $\mbox{dist}(p_G,p_*)$ denotes the distance between two probability distributions such as the Pearson $\chi^2$ divergence $\mbox{dist}(p_G,p_*)=\chi^2_{Pearson}(p_G+p_*,2p_*)$. Here, $E[\z]$ is the expectation of $\z$ and $\lambda>0$ is the regularization parameter. Minimizing the model in \eqref{model0} with $\mbox{dist}(p_G,p_*)$ using the Pearson $\chi^2$ divergence can be converted to the least squares generative adversarial networks framework \cite{Mao2019}; $G$ is trained simultaneously with a discriminator $D$ in an adversarial relationship to improve their mutual abilities as follows:
\begin{align}\label{gu}
\left\{ \begin{array}{l}
\W_g^* =  {\displaystyle \arg\min_{\W_g}} ~~  E_{\z\sim p_{\mbox{\tiny FDK}}} \left[ D(G(\z;\W_g))^2\right] + \lambda  E_{\z\sim p_{\mbox{\tiny FDK}}} \left[\|G(\z;\W_g)-\z\|_2^2 \right] \\
\W_d^*={\displaystyle \arg\min_{\W_d}} ~~  E_{\mmu\sim p_{*}}\left[(1-D(\mmu;\W_d))^2\right] + E_{\z\sim p_{\mbox{\tiny FDK}}}\left[(1+D(G(\z);\W_d))^2 \right].
\end{array}
\right.
\end{align}
A detailed derivation of \eref{gu} is described in \cite{Mao2019,Park2020}. The f-divergence, including the Pearson $\chi^2$ divergence, is not symmetric, and therefore does not satisfy the metric properties \cite{Nowozin2016}. Some studies \cite{Arjovsky2017,Lee2021} have exploited the Wasserstein distance to satisfy metric properties in order to match the probability distribution $p_{*}$.

\begin{figure}[ht]
	\centering
	\includegraphics[width=1\textwidth]{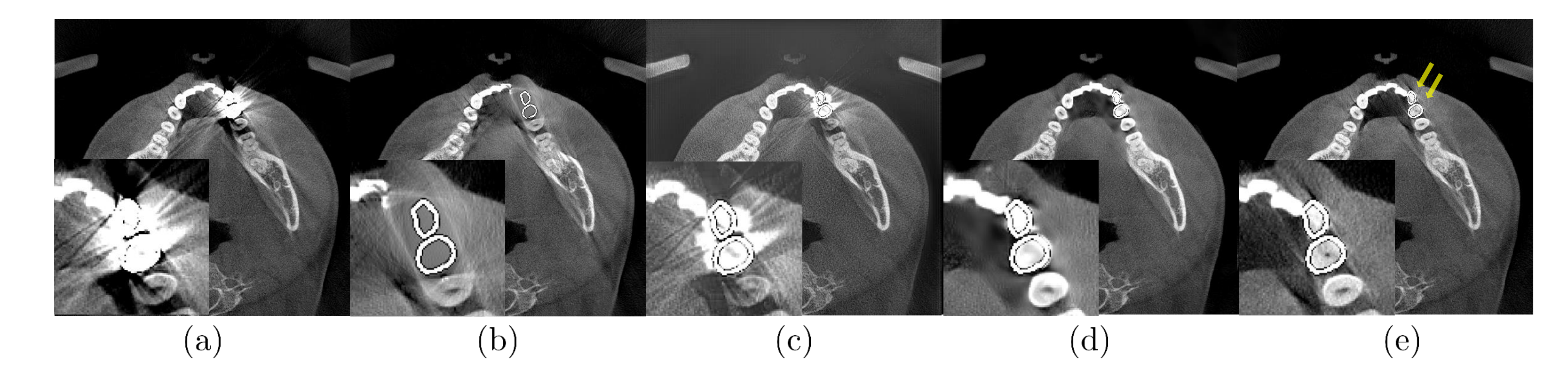}
	\caption{Visual comparison of  MAR methods for realistic test dataset. The figure in (a) shows clinical dental CBCT image $(\mathcal C^\dagger_{\mbox{\tiny FDK}}[\mathbf{P}])$ with two simulated dental crowns (see yellow arrows in (b)). The severe metal artifacts owing to the two crowns occur in $\mathcal C^\dagger_{\mbox{\tiny FDK}}[\mathbf{P}]$. The figures in (b), (c), and (d) show corrected image obtained by NMAR \cite{Meyer2010}, LSGAN  \cite{Mao2019}, the data-driven regularizaion approach \cite{Park2022}, respectively. The figure in (e) shows the corresponding target image $(\mmu^*)$. }
	\label{fig:results_DL}
\end{figure}

If the gap between $\z$ and the corresponding artifact-free image is not small, the fidelity $E_{\z\sim p_{\mbox{\tiny FDK}}} \left[\|G(\z;\W_g)-\z\|_2^2 \right]$ may not be sufficient for MAR. Even when using deep learning techniques in a learning environment where only unpaired training data are available, it can be very difficult to accurately reconstruct the details of the tooth surface from projection data $\mathbf{P}$  heavily contaminated by metal implants, as shown in Fig. \ref{fig:results_DL}. In particular, the structures of the teeth and the oral cavity differ from person to person. It seems to be necessary to secure and supplement sophisticated tooth information to obtain an image reconstruction accurate enough for dental prosthetic treatment.

We note that the 3D segmentation of teeth, jaw, and skull from a CBCT image is an important component of 3D cephalometric analysis, where soft tissue details are not required \cite{Yun2022}. Therefore, dental CBCT reconstruction can focus on restoring the morphological structures of the bones and teeth near metal objects instead of considering soft tissues. This simplified approach allows the use of supplementary techniques to obtain an image before dealing with the ill-posed problem.

However, in the case of CBCT images deteriorated by metal artifacts, the boundaries between teeth are often occluded, making it difficult to accurately segment a 3D individual tooth. Recently, several deep learning methods \cite{lee2020automated,rao2020symmetric,cui2019toothnet} have been proposed for automated 3D tooth segmentation directly from $\mu$; however, their performance is far from satisfactory on CBCT images with severe metal artifacts.

\subsection{2D panoramic image prior for 3D individual tooth segmentation}
\begin{figure}[h]
	\centering
	\includegraphics[width=0.95\textwidth]{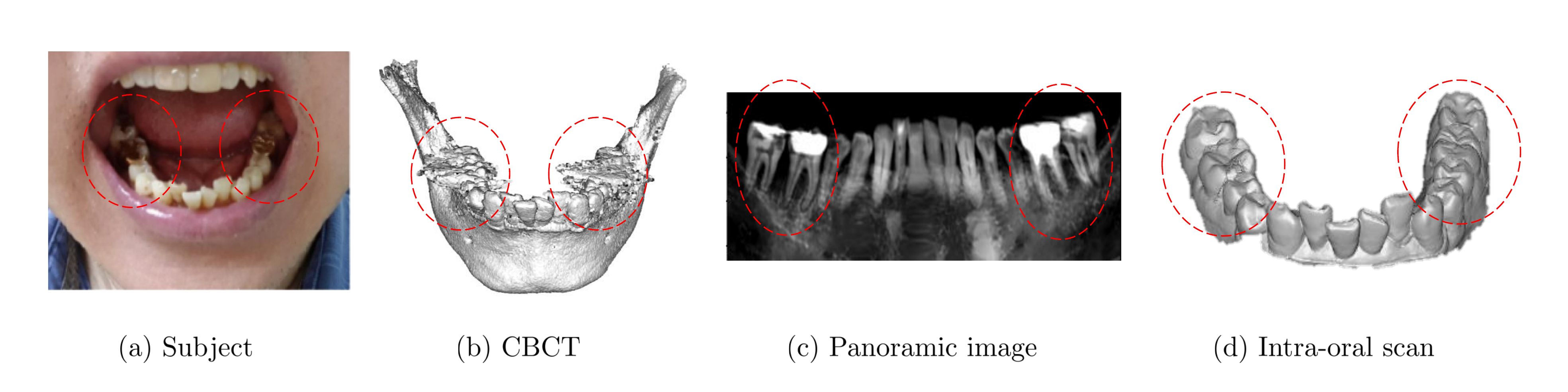}
	\caption{Concrete image priors; panoramic image and intra-oral scan.}
	\label{fig:panorama}
\end{figure}

Jang \etal \cite{Jang2021} observed that panoramic images generated from CBCT images were not significantly affected by metal-related artifacts. This is because the cone beam projection configuration is advantageous for composing panoramic image reconstructions. Fig. \ref{fig:panorama} (b) shows a 3D bone-teeth-jaw model for one of the authors who has multiple gold dental prostheses (Fig. \ref{fig:panorama} (a)). The 3D bone-teeth-jaw model in Fig. \ref{fig:panorama} (b) was generated from the CBCT image using state-of-the-art MAR software. However, state-of-the-art MAR does not adequately remove metal artifacts caused by the crown and does not fully recover the neighboring tooth details. The panoramic image generated by CBCT data (Fig. \ref{fig:panorama} (c)) produce effective prior information. Jang \etal \cite{Jang2021} leveraged these panoramic images as a prior to accurately perform 3D tooth segmentation and identification.

\begin{figure}[h]
	\centering
	\includegraphics[width=0.95\textwidth]{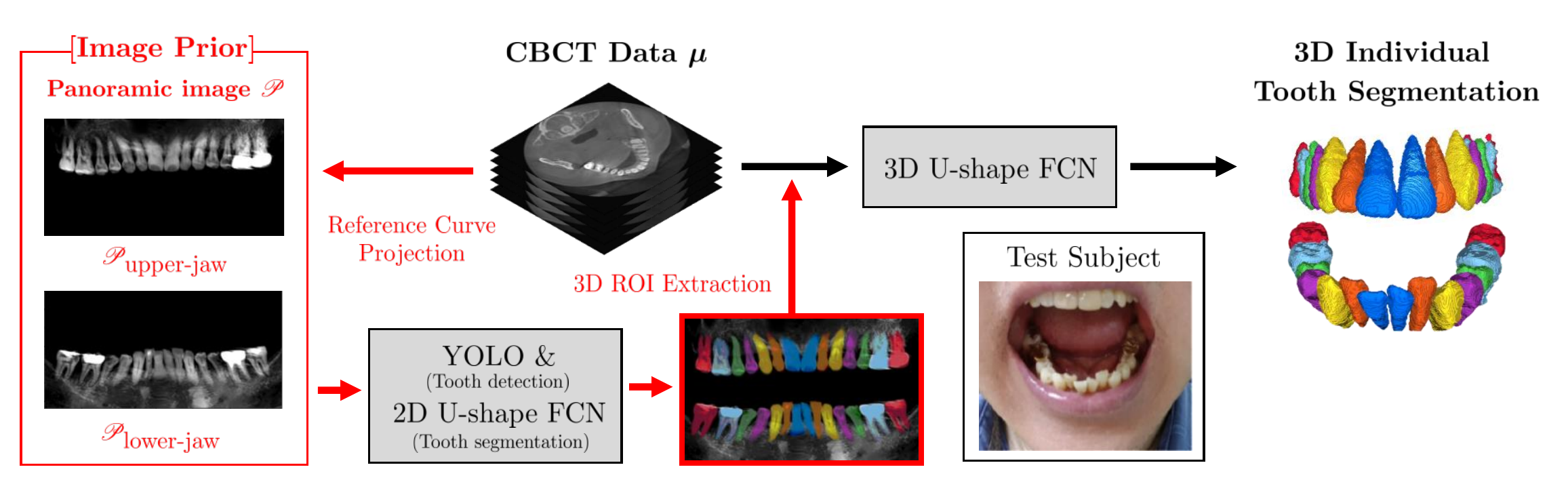}
	\caption{3D individual tooth segmentation method \cite{Jang2021} using panoramic images. The figure above shows overall processes and results in the test case of a subject having metal prostheses.}
	\label{ToothSegAlgorithm}
\end{figure}

Let us briefly explain how to automatically generate upper and lower jaw panoramic images from a 3D CBCT image \cite{Jang2021}. The upper and lower jaws are separated to reduce the overlap between adjacent teeth. To explain the panoramic image generation, we used the CBCT image, which is seriously degraded by metal artifacts.

Generating panoramic images begins by obtaining binary images (or segmentations) of the upper and lower jaws by applying Otsu's thresholding technique \cite{Otsu1979} and connected-component labeling \cite{Samet1988}.  Using the upper-jaw binary image, we obtain the following 2D reference curve $\mathcal{C}_{\mbox{\tiny upper-jaw}}$ passing through the upper dental arch region completely:
\begin{equation}	
	\mathcal{C}_{\mbox{\tiny upper-jaw}} = \{ \x_u(s): s \in [0,1] \}.
	\label{eq:curve}
\end{equation}
Similarly, we can obtain $\mathcal{C}_{\mbox{\tiny lower-jaw}}$ from the lower dental arch region. We then project the CBCT image $\mu$ along the curve normal direction to generate the upper jaw panoramic image $\mathscr P_{\mbox{\tiny upper-jaw}}$, which is given by
\begin{equation}
	\mathscr P_{\mbox{\tiny upper-jaw}}(s,z) = \int^{a}_{-a} \mu_{\mbox{\tiny upper-jaw}} \left( \x_u(s)+t\bold{n}(s),z \right)\,dt,
	\label{eq:pan}
\end{equation}
where $\mu_{\mbox{\tiny upper-jaw}}$ is the upper-jaw CBCT image, $s$ is the parameter in \eqref{eq:curve} and $\bold{n}(s)$ is the unit normal vector at $\x_u(s)$. Similarly, we obtain the lower panoramic image $\mathscr P_{\mbox{\tiny lower-jaw}}$, as shown in Fig. \ref{ToothSegAlgorithm}.

We next use a U-shaped fully-convolutional network \cite{ronneberger2015u} and YOLO \cite{redmon2016you} to obtain the 2D tooth segmentation, as shown in Fig. \ref{ToothSegAlgorithm}. Next, backprojecting this 2D individual tooth segmentation along reference curve $\mathcal{C}_{\mbox{\tiny lower-jaw}}$  provides a tight 3D ROI of the corresponding individual tooth in the 3D CBCT image $\mu$. These tight ROIs on individual teeth are crucial for fine 3D individual tooth segmentation at the boundary where the target tooth and adjacent teeth meet \cite{Jang2021}. Therefore, the panoramic image, as an image prior, plays an important role in 3D tooth segmentation.

\subsection{Using radiation-free intra-oral scan data to obtain teeth surface priors}
Although the method using the panoramic image prior obtains accurate tooth segmentation in the case of CBCT images severely affected by metal artifacts, it still has limitations in accurately restoring tooth surfaces. It seems to be fundamentally difficult to precisely restore tooth surfaces around metal implants on CBCT images that are severely damaged by metal-induced artifacts.

The use of an intra-oral scan (IOS) as a concrete image prior can effectively compensate for the aforementioned weakness of dental CBCT without increasing X-ray dose exposure. Recently, intra-oral scanner equipped with cutting-edge technology was developed that can acquire accurate 3D images of the teeth surfaces and gingiva at high resolution \cite{zimmermann2015intraoral}, and its accuracy is approaching the level required for clinical application for digital impressions and occlusal analysis \cite{zarone2020accuracy}. Fig. \ref{fig:panorama} (d) shows that the IOS provides precise tooth surface images, whereas dental CBCT images can be affected by metal-related artifacts.

Hyun \etal \cite{Hyun2022} leveraged tooth surface information from the IOS to compensate for the damage of CBCT images caused by metal-induced artifacts. By merging the IOS into CBCT scans via a surface matching method \cite{Jang2022}, they provided an accurate jaw-teeth model for a realistic digital simulation. This method can facilitate the virtual surgical planning, treatment simulation, and design and delivery of orthodontic and surgical treatment \cite{elnagar2020digital, baan2021fusion, li2021creating}.

\begin{figure}[h]
	\centering
	\includegraphics[width=0.95\textwidth]{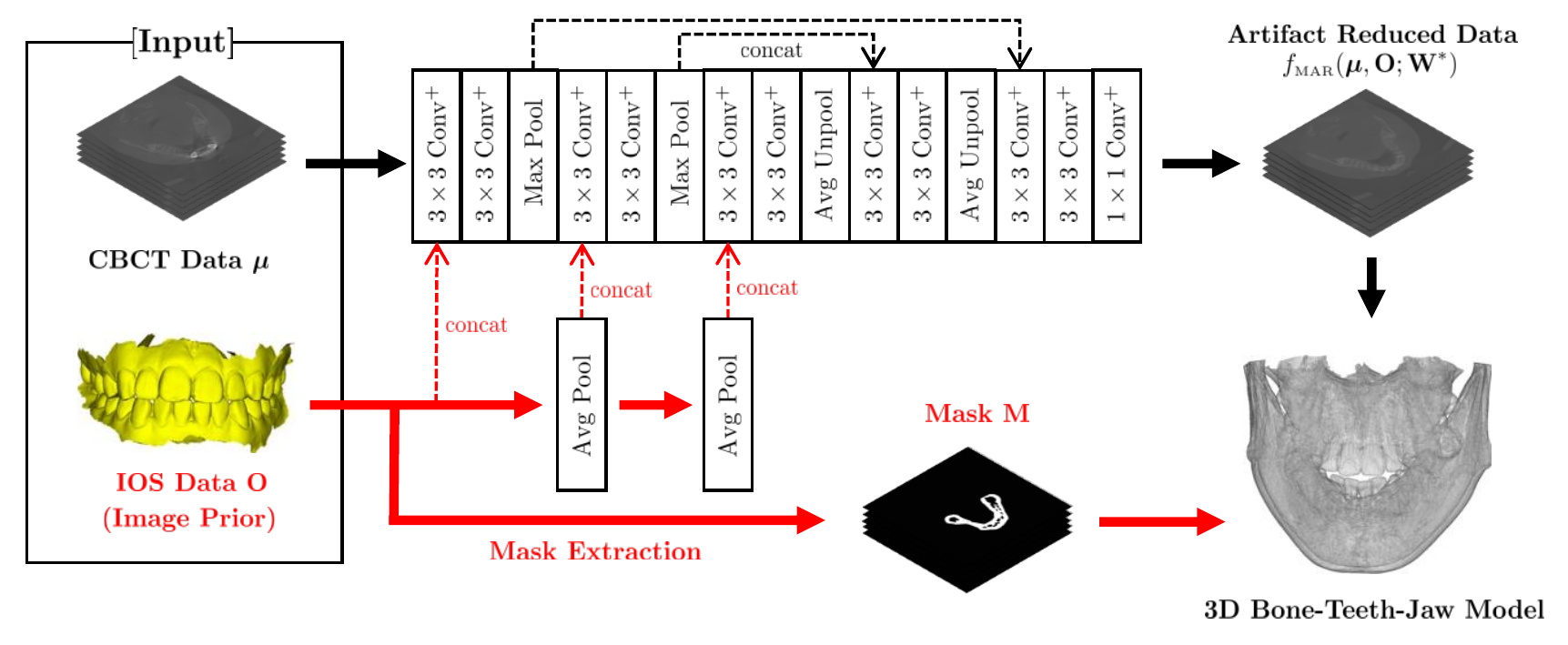}
	\caption{Schematic diagram of the method for 3D bone-teeth-jaw modeling using IOS data $\textbf{O}$ \cite{Hyun2022}. The IOS data $\textbf{O}$ is utilized as prior information of teeth surface for metal artifact reduction and 3D modeling.}
	\label{MARIOSalgorithm}
\end{figure}

Fig. \ref{MARIOSalgorithm} shows the process of using the prior information about tooth surface obtained from IOS. Let $\mathbf O$ represent the IOS data. Given data triplets $\{(\mmu_i,\mathbf O_i, \mmu^*_i)\}_{i=1}^{L}$, a MAR network $f_{\text{MAR}} : (\boldsymbol \mu,\mathbf O) \mapsto \boldsymbol \mu^*$ is trained by solving the following minimization problem:
\begin{equation} \label{IOSMAR}
	\W_* = \underset{\W}{\mbox{argmin}} ~ \dfrac{1}{L} \sum_{i=1}^{L} \| f_{\text{MAR}}(\boldsymbol \mu_i,\mathbf O_i;\W) -  \boldsymbol \mu^*_i \|^2_2,
\end{equation}
where $\mathbf O_i$ is used as a side input in $f_{\text{MAR}}$. As shown in Fig. \ref{MARIOSalgorithm}, the IOS data can also provide a mask $\mathbf M$ that does not contain teeth, and $\mathbf M$ is applied to the corrected CBCT image $f_{\text{MAR}}(\boldsymbol \mu,\mathbf O;\W_*)$ to obtain a 3D bone-teeth-jaw model.  Fig. \ref{MARIOS} (c) shows the 3D bone-teeth-jaw model constructed using the two-stage method.

\begin{figure}[h]
	\centering
	\includegraphics[width=0.8\textwidth]{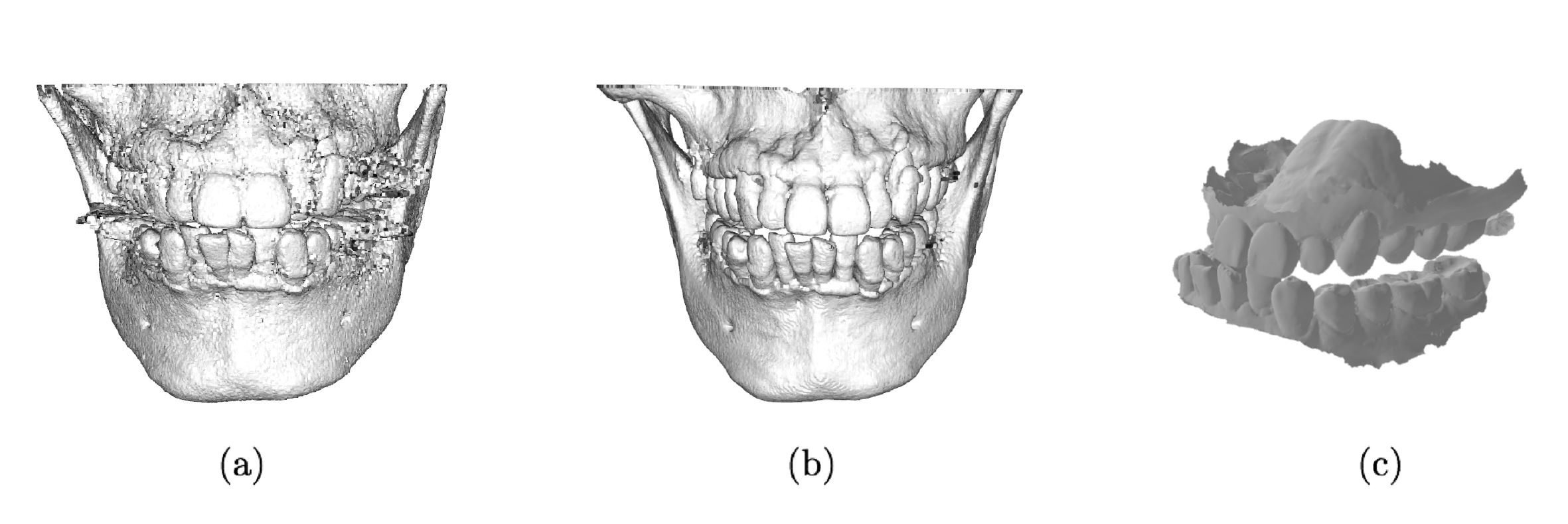}
	\caption{CBCT-based 3D bone-teeth-jaw modeling. (a) shows 3D bone-teeth-jaw model generated from uncorrected CBCT image with the simple thresholding. The figure in (b) shows 3D bone-teeth-jaw model obtained from the method \cite{Hyun2022} that leverages the tooth surface information from IOS in (c).}
	\label{MARIOS}
\end{figure}
\section{Discussion and conclusion}
Dental CBCT aims to provide high-resolution images with the lowest possible radiation dose at a low cost for equipment and maintenance. This cost-competitive goal of dental CBCT makes the data acquisition hardware configuration different from that of MDCT, which is widely used in clinical practice. As a result, the inverse problem in dental CBCT is more ill-posed than it is in MDCT.

The fundamental reasons why dental CBCT is more ill-posed than MDCT are as follows. The data acquisition of MDCT is based on helical CT scanning and continuous table movement, and the fastest rotation time of MDCT is 0.33 seconds. By contrast, a dental CBCT system uses a fixed array of detectors and the body is scanned in one revolution. Because the rotation time of dental CBCT is more than 8 s, motion artifacts may also occur. Most dental CBCT devices use an offset detector with a short subject-to-detector distance (or air gap) to obtain a larger FOV with a small detector as possible. Because of the short subject-to-detector distance, the most serious source of artifacts in dental CBCT is scattering. As the air gap decreases, the probability that the scattered photons escape out of the detector decreases \cite{Smith1985,Neitzel1992}. This scattering effect makes the forward model \eqref{Beerslaw-CBCT} less accurate.

Despite the disadvantages mentioned above, dental CBCT systems are rapidly growing in demand owing to their cost competitiveness and low radiation dose, enhancing the confidence of clinicians who operate the equipment. Metal artifacts are common in dental CT. As the number of older people with artificial prostheses and metallic implants rapidly increases as the population ages, it is very important to deal with the inverse problem in which data are damaged by metal objects. Despite various studies seeking to reduce metal artifacts, metal streaking artifacts continue to pose difficulties, and the development of suitable reduction methods remains challenging.

Recently, numerous attempts have been made to use deep learning for MAR \cite{Zhang2018-1,Park2018,Gjesteby2019,Lin2019,Yu2020}. These deep learning-based MARs have demonstrated remarkable performance in limited environments. However, their performance in dental CBCT environments is limited when multiple metallic inserts occupy a significant area. It seems that there is a fundamental limitation in the accuracy of morphological tooth structure restoration when only dental CBCT data severely damaged by metal implants are used. Therefore, it would be desirable to use radiation-free intraoral scan data with dental morphological structures as prior information in image reconstruction. Our current research topic is the development of a deep learning method that effectively uses the tooth shape information obtained from an oral scanner for CBCT image reconstruction.

The development of artificial intelligence is expected to automate the convergence of CBCT, oral scanners, and facial scanners, which will substantially help both patients and doctors manage dental care and dental health. The integration of CBCT and IOS can provide highly accurate digital impressions by compensating for the shortcomings of metal artifacts in CBCT. Traditional impression-making methods have a number of factors that limit their accuracy, such as patient movement, tearing and deformation of the impression during removal, and soft tissue contraction. Therefore, this fusion approach could eliminate the cumbersome procedure of traditional impressions for both the dentist and patient, significantly shortening the treatment time.

\section*{Acknowledgment}
This work was supported by Samsung Science $\&$ Technology Foundation (No. SRFC-IT1902-09). H S Park was partially supported by the National Institute for Mathematical Sciences(NIMS) grant funded by the Korean government (No. NIMS-B22920000).

\end{document}